# Sound insulation performance of multi-layer membrane-type acoustic metamaterials based on orthogonal experiments

Jun Lan,[1] Yumin Zhou,[1] Xin Bu,[1] and Yifeng Li,[1,2,a]

[1]College of Computer and Information Engineering (College of Artificial Intelligence), Nanjing Tech University, Nanjing 211800, China

[2]Key Laboratory of Modern Acoustics, Institute of Acoustics and School of Physics, Collaborative Innovation Center of Advanced Microstructures, Nanjing University, Nanjing 210093, China

[a]Author to whom correspondence should be addressed: lyffz4637@163.com

**ABSTRACT**

The challenge of achieving effective sound insulation using metamaterials persists in the field. In this research endeavor, a novel three-layer membrane-type acoustic metamaterial is introduced as a potential solution. Through the application of orthogonal experiments, remarkable sound insulation capabilities are demonstrated within the frequency spectrum of 100-1200 Hz. The sound insulation principle of membrane-type acoustic metamaterial is obtained through the analysis of eigenmodes at the peak and trough points, combined with sound transmission loss. In addition, an orthogonal experiment is utilized to pinpoint the critical factors that impact sound insulation performance. By using relative bandwidth as the classification criterion, the optimal combination of influencing factors is determined, thereby improving the sound transmission loss of the multi-layer membrane-type acoustic metamaterial structure and broadening the sound insulation bandwidth. This study not only contributes a fresh and practical approach to insulation material design but also offers valuable insights into advancing sound insulation technology.

## 1. Introduction

Noise pollution is a serious issue in modern society, negatively impacting health and daily life. In particular, mid-to-low frequency noise affects attention and emotion, leading to psychological issue[1]. Acoustic metamaterials[2] offer a new solution to these problems. Currently, acoustic metamaterials employ various innovative structures in sound insulation to achieve efficient isolation and absorption of different frequency sound waves[3]. These designs include phononic crystal structures[4-7], local resonance structures[8-11], micro-perforated structures[12-14], etc. Phononic crystal structures are suitable for controlling sound waves within specific frequency ranges, but the manufacturing process is complex and lack adaptability. Localized resonance structures are widely used in isolating low-frequency sound waves and are particularly suitable for applications that require efficient and compact designs. However, achieving such performance often necessitates increased mass and meticulous engineering. Micro-perforated structures can effectively absorb and attenuate mid-to-high frequency sound waves, while low-frequency absorption capabilities are limited. These structures have their own unique working principles and application domains, providing a wide range of choices to cater to different sound absorption needs.

Recently, Membrane-type acoustic metamaterials (MAM)[15-19] with negative mass density have been proposed, either as individuals or as ensembles as ideal candidates for low-frequency sound insulation in acoustic devices. Compare with single MAM, the integration of multi-unit becomes imperative to attain improved performance and broader applicability. The design of multi-unit plays a pivotal role in optimizing their effectiveness and expanding their scope of use. For example, Zhou et al[20]. introduced a low-frequency broadband MAM with multiple resonant modes, expanding low-frequency sound attenuation range and demonstrating single-negative effective parameters. Jang et al. presented a lightweight soundproofing MAM structure, achieving significant broadband soundproofing performance improvement by utilizing thin film vibration[21]. Ciaburro, G et al. proposed a three-layer MAM absorber structure using recycled materials, enhancing sound absorption performance in the low-frequency range and delivering exceptional results across a wide frequency range[22]. Recently, Li et al. designed a sandwich structure combining a dual MAM with Helmholtz resonators, exhibiting superior sound insulation performance and allowing for flexible adjustment to meet diverse engineering requirements[23]. These studies indicate that the multi-unit MAM exhibits superior sound insulation performance compared to single-cell ones[24].

The purpose of this work is to explore the combination of MAM units, from single-unit to multiple-unit structures. By studying the multi-layer structure of MAM, we can gain insights into its acoustic characteristics and optimization methods[25]. In this study, we establish a three-layer structure based on existing polymorphic anti-resonant cooperative MAM to investigate its sound insulation performance. The influence of the MAM multi-layer structure and array arrangement on sound insulation is examined through theoretical analysis, numerical simulation and simulation experiment. The sound insulation performance of the multi-layer MAM structure is optimized through orthogonal experiments. These findings provide valuable references for optimizing MAM design and its applications, promoting the field of noise reduction and beyond.

## 2. Model introduction and theoretical analysis

### 2.1 Structure design

A typical MAM cell comprises a polyimide (PI) membrane for elasticity, an ethylene-vinyl acetate copolymer (EVA) rigid frame for securing the membrane, and resonators made of dense metal masses. The entire MAM vibration system can be effectively equivalent to a "spring-mass" system. By arranging the resonators and membrane rationally, excellent sound insulation performance can be achieved. As shown in Fig. 1, the metal resonators and EVA swing arms are placed on the same side of the PI membrane, which is fixed within the EVA ring frame[26]. The concentric radial structure of the spider web structure provides excellent rigid support through hierarchical load distribution, offering significant advantages in terms of lightweight design compared to traditional lattice structures. Furthermore, the spider web structure is well-suited for efficient production using modern manufacturing techniques and features flexible design, allowing for parameter adjustments to optimize performance according to specific requirements[27,28]. To match the characteristic of a spider web, the membrane is simulated as a spider web structure. In the MAM cell, the cross-shaped swing arm is positioned at the center of the membrane, with rectangular resonators attached radially to approximate the radial spider web shape. The resonators are attached to the nodes of the PI membrane, forming an overall circumferential arrangement. The cross-shaped swing arm divides the membrane into four sub-regions, and the metal resonators further divide the sub-regions, creating three membrane regions for the resonators, swing arm and swing arm tail end, respectively. The spider web-inspired structure divides the MAM cell into twelve symmetrically distributed sub-cells, with alternating material parameters for each sub-cell. The material parameters of the MAM cell are presented in Table 1.

**Table 1.** The material parameters of the MAM cell

| Parameter | Thickness (mm) | Diameter (mm) | Width (mm) | Length (mm) | Modulus (Pa) | Density (kg/m$^3$) | Poisson's ratio |
|---|---|---|---|---|---|---|---|
| Frame | 2 | 50 | 5 | | $1.70\times10^8$ | 2050 | 0.45 |
| Swing-arm | 2 | | 4 | 40 | $1.70\times10^8$ | 2050 | 0.45 |
| Membrane | 0.2 | 50 | | | $1.42\times10^9$ | 1100 | 0.36 |
| Resonator | 1.8 | | 4 | 13 | $2.00\times10^{11}$ | 7800 | 0.33 |

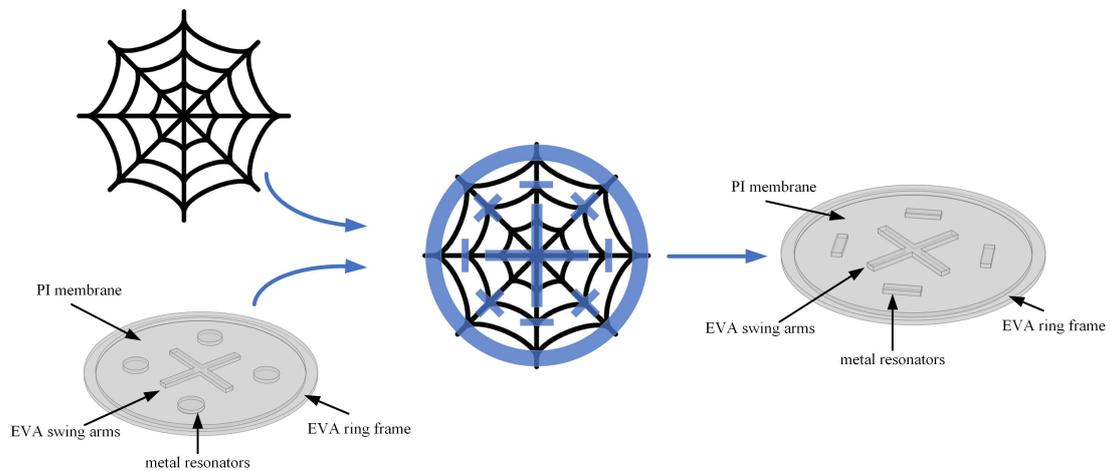

**Fig. 1.** Simplified MAM design based on spider web topology.

## 2.2 Simulation model and theoretical analysis

Based on the COMSOL Multiphysics software, a three-layer MAM finite element simulation model is constructed to investigate its sound insulation performance within a frequency range of 50-1000 Hz with a step of 10 Hz. As shown in Fig. 2(a), based on the designed MAM cell, a tubular EVA frame is used, and the membrane is fixed inside the tube. The resonators and cross-shaped swing arms are attached to the same side of the PI membrane. In this configuration, the entire structure can be seen as three MAM cells with two cavities sandwiched between the MAM cells. The width of the cavities ($H_c$) is set to 25 mm, and the overall height of the structure is 65 mm. All other parameters remain consistent with the MAM cell parameters. This simulation model includes both the solid domain and the pressure acoustics domain. In the solid domain, the frame and the membrane are surrounded by fixed boundaries to simulate the overall sound insulation performance under fixed constraints, with an initial prestress of 1 MPa in the membrane. The loss factor for the PI film is calculated for enhances the accuracy. The two ends of the pressure acoustics domain are set as plane wave radiations. The upper face serves as the sound wave entrance, with a vertically incident plane wave amplitude of 1 Pa. The lower face serves as the sound wave exit, with a non-reflecting boundary. The density of the air medium is 1.29 kg/m$^3$, and the speed of sound in air is 343 m/s.

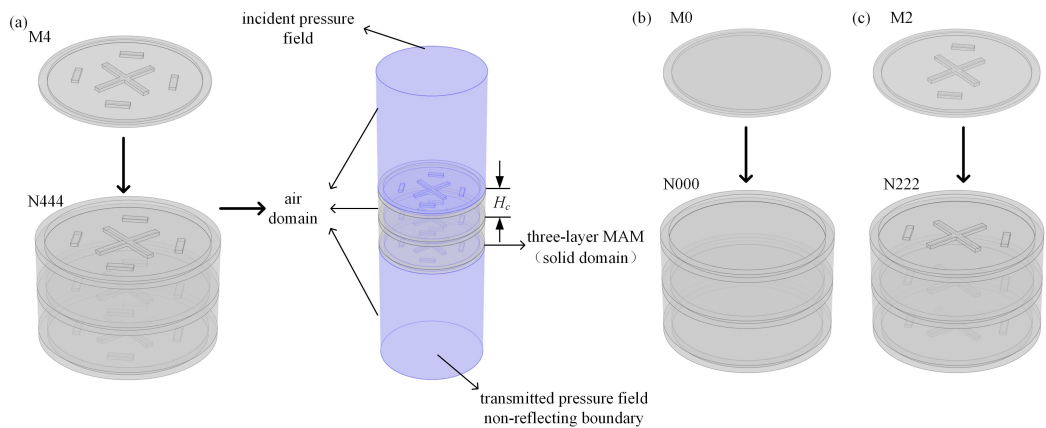

**Fig. 2.** The simulated structures. (a) M4 and N444; (b) M0 and N000; (c) M2 and N222.

The mass law and sound transmission loss (STL) serve as theoretical foundations and important indicators for evaluating the sound insulation performance of materials[29]. STL refers to the amount of energy attenuation of sound waves when passing through structure, and it is a crucial parameter for measuring sound insulation performance. STL can be calculated using the transmission coefficient $t_I$:

$$\text{STL}=10\lg\frac{1}{t_I} \tag{1}$$

First, the sound pressures in the incident pressure field $P_{in}$ and the transmitted pressure field $P_{out}$ are integrated to calculate the incident sound power $W_{in}$ and the transmitted sound power $W_{out}$, respectively, which can be expressed as:

$$W_{in} = \int (P_{in}^2 / 2\rho_a c_a) dS_{in} \tag{2}$$

$$W_{out} = \int (P_{out}^2 / 2\rho_a c_a) dS_{out} \tag{3}$$

where $\rho_a$ and $c_a$ represent the density and speed of sound in air, and $S_{in}$ and $S_{out}$ represent the areas of the incident pressure field and transmitted pressure field surfaces, respectively. Hence, STL can be expressed as follows:

$$\text{STL}=10\lg(W_{in} / W_{out}) \tag{4}$$

The structure is composed of multiple different materials, and there are cavities between the layers, where the acoustic impedance of air differs significantly from the simplified structure. In theoretical analysis, the characteristic impedance of the structure is assumed to be constant, allowing for simplification in the analysis. When sound waves are incident, the theoretical simplified model is shown in Fig. 3(a), while the actual transmission mode of the model is illustrated in Fig. 3(b). Media I and III represent air, while Medium II represents the effective structure. The expression for sound wave transmission through the structure is as follows:

$$\begin{aligned} P_{in} &= P_{ia} e^{j(\omega t - k_1 x)} \\ P_{1r} &= P_{1ra} e^{j(\omega t + k_1 x)} \\ P_{2t} &= P_{2ta} e^{j(\omega t - k_2 x)} \\ P_{2r} &= P_{2ra} e^{j(\omega t + k_2 x)} \\ P_{out} &= P_{ta} e^{j[\omega t - k_1(x-D)]} \end{aligned} \tag{5}$$

Where $\omega$ is the angular frequency, and $D$ represents the thickness of medium II,

which is the total thickness of the three-layer MAM structure.

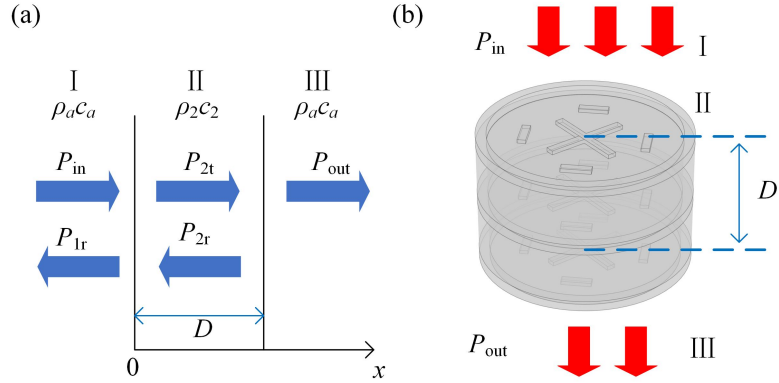

**Fig. 3.** Propagation model. (a) the theoretical simplified model; (b) the actual structural model.

## 3. Sound insulation performance

### 3.1 Sound insulation characteristics

STL is used as an evaluation standard for sound insulation performance. The three-layer MAM model is defined based on the number of resonators in each layer. Figure 4(a) demonstrates the STL of the N444 model. N444 has a sound insulation bandwidth of 120-860 Hz, with a peak STL of 104 dB at 460 Hz. Compared to the single layer MAM cell M4, N444 exhibits significant improvements in low-frequency sound insulation within the 330-810 Hz range. The overall trend of the N444 STL remains consistent with the M4, with minimal differences in valley and peak frequencies and sound insulation bandwidth. However, sound waves passing through the three-layer MAM of N444, especially at the peak frequency, the sound energy undergoes effective cancellation and transfer, resulting in a multiplied sound insulation peak.

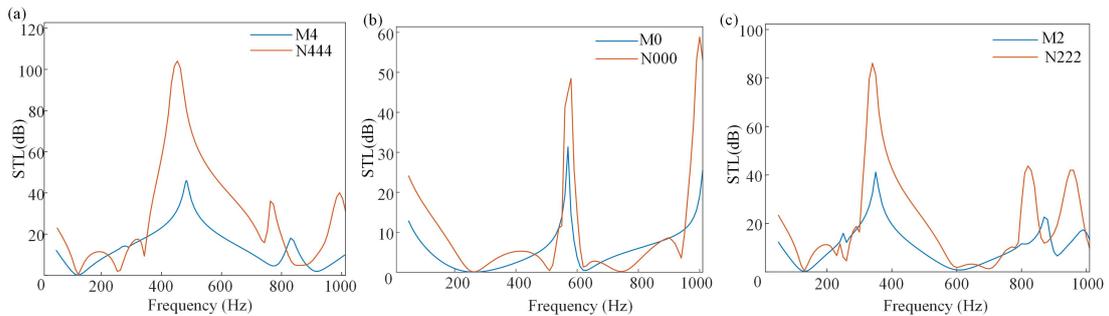

**Fig. 4.** Comparison of the STLs for single-layer and three-layer MAMs. (a) M4 and N444; (b) M0

and N000; (c) M2 and N222.

To achieve a lightweight design and generate more anti-resonance modes, the first step is to explore the reduction of resonator quantity. The mass of the resonators directly impacts the equivalent mass of the MAM vibration system, thereby influencing its natural frequencies. By decreasing the number of resonators, we can generate anti-resonance modes at different frequencies. To maintain symmetrically opposite vibrational displacements in the membrane, the number of resonators can only be 0, 2 or 4. Accordingly, the single layer MAM cells are defined as M0, M2 and M4, respectively. Figures 2(b) and 2(c) present the structural diagrams of M0 and M2, as well as their three-layer configurations N000 and N222. Aside from the variation in the number of resonators, all other parameters remain consistent with M4 and N444. STLs of the corresponding models are depicted in Figs. 4(b) and 4(c), respectively. The sound insulation bandwidth of N000 is 260-620 Hz, with a peak STL of 48.5 dB at the frequency of 580 Hz. N222 has a sound insulation bandwidth of 130-600 Hz, with a peak STL of 86.1 dB at the frequency of 340 Hz. It is observed that increasing the number of resonators widens the sound insulation bandwidth and improves the peak of the STL. However, similar to the comparison with M4 and N444, in N000 and N222, the peak STL at the frequency point doubles, while the sound insulation bandwidth remains essentially unchanged, and no new anti-resonance modes are generated. Therefore, to widen the sound insulation bandwidth of the three-layer MAM structure, it is not simply a matter of increasing the number of resonators, but also considering the structure and arrangement of each layer's MAM cells.

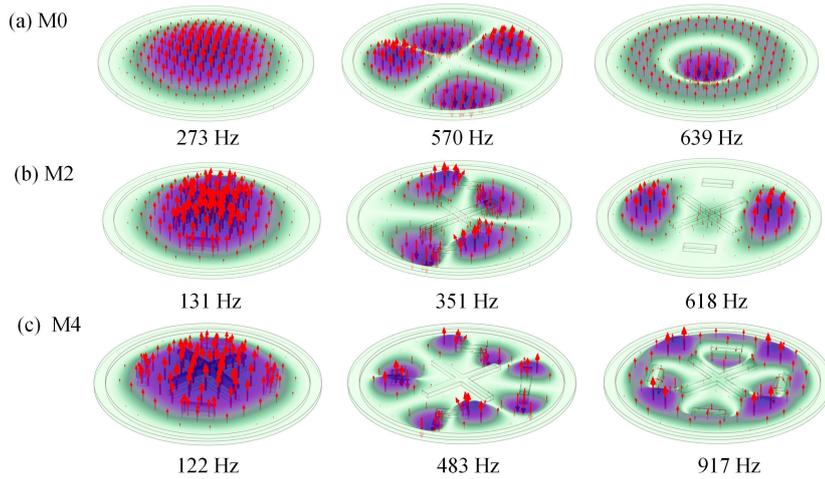

**Fig. 5.** Eigenmode contours. (a) M0; (b) M2; (c) M4.

In the following, the eigenmode patterns of the peak and trough points within the bandwidths of M0, M2 and M4 are analyzed in conjunction with the STLs to understand the sound insulation mechanism. The eigenmode contours of M0, M2 and M4 are shown in Figs. 5(a), 5(b) and 5(c), respectively. The membrane vibrates in the same direction as the incident sound wave at 273 Hz for M0, and the sound energy is not cancelled by any reverse sound wave, which resulting in the maximum sound transmission. The resonator and swing-arm vibrate in phase with the membrane at 131 Hz for M2 and 122 Hz for M4, forming an overall coupled resonant mode, and the vibration direction aligns with the incident sound wave, resulting in the maximum sound transmission. The membrane exhibits symmetrical antiphase vibration displacements near 570 Hz for M0, and the incident sound energy is cancelled by the local resonance of the membrane, resulting in the maximum STL value. The membrane near the resonators exhibits symmetrical antiphase vibration displacements at 351 Hz for M2 and 483 Hz for M4, and the entire cell is in a quasi-dynamic equilibrium state, resulting in the maximum STL value. The majority of the membrane's vibration direction aligns with the incident sound wave at 639 Hz for M0, but a small portion of the reverse sound wave cancels out the sound energy, forming a trough. The membrane regions without resonators undergo in-phase vibrations at 618 Hz for M2 and 917 Hz for M4, while the resonators have smaller vibration amplitudes. As a result, a large amount of sound energy is gathered and transmitted through the membrane regions, forming a trough.

### 3.2 Sound insulation performance factors

By conducting an analysis of the soundproof performance of the three-layer MAM, it is necessary to explore the impact of changes in each layer's structure on the overall soundproof performance. All the designed simulation models are shown in Fig. 6(a). Based on the N000 structure, each layer is enhanced with an M4 structure, which includes four resonators and swing arms. The changes in STL are investigated, as shown in Fig. 6(b). Compared to N000, N004 exhibits a new STL peak in the frequency range of 200-550 Hz, with a peak value of 37.7 dB at the frequency of 460 Hz. This peak corresponds to the resonant frequency of the M4 structure, indicating that it originates from the anti-resonance behavior of the M4 structure. Compared to N004, N044 exhibits a peak STL of 68.4 dB in the frequency range of 170-560 Hz, with the frequency of 450 Hz. This confirms the emergence of new anti-resonance modes in N044. The generation of the two STL peaks is attributed to the anti-resonance behavior of M0 and M4 structures, respectively. The overall STL is enhanced under the coupled vibration of these two structures. On the basis of N044, a comparison is made with N022 and N024 to investigate their STL variations, as shown in Fig. 6(c). N022 exhibits a new STL peak in the frequency range of 280-480 Hz, with a peak value of 59.1 dB at the frequency of 350 Hz. This peak corresponds to the resonant frequency of the M2 structure. N024 exhibits multiple anti-resonance modes in the frequency range of 300-800 Hz, with three STL peaks. Compared to N022, N024 demonstrates superior soundproof performance after 380 Hz. Compared to N044, N024 shows improved soundproof performance before 380 Hz. Although the soundproof peak of N024 at 380 Hz is not as high as N044, the soundproof level of N024 still meets the application requirements, and it is also lighter in weight. Therefore, with the goal of generating more anti-resonance modes and ensuring lightweight design, the N024 structure is considered the optimal choice. The STL of the three-layer MAM is the result of the coupling of different structures. The order of the layers does not affect the overall soundproof performance.

In the above study, we discussed the impact of the structure and the order of each layer on the soundproof performance. The overall structure not only includes MAM cells but also cavities located within the two-layer structure. To investigate the influence of the cavity structure on the overall soundproof performance, cavity thicknesses of 15 mm, 25 mm and 35 mm are chosen. The results are shown in Fig. 6(d). It is observed that, when the cavity thickness increases, the soundproof performance in the entire frequency range improves, while the overall bandwidth remains relatively unchanged. Therefore, increasing the cavity thickness can further enhance the dissipation of sound energy and improve the overall soundproof performance.

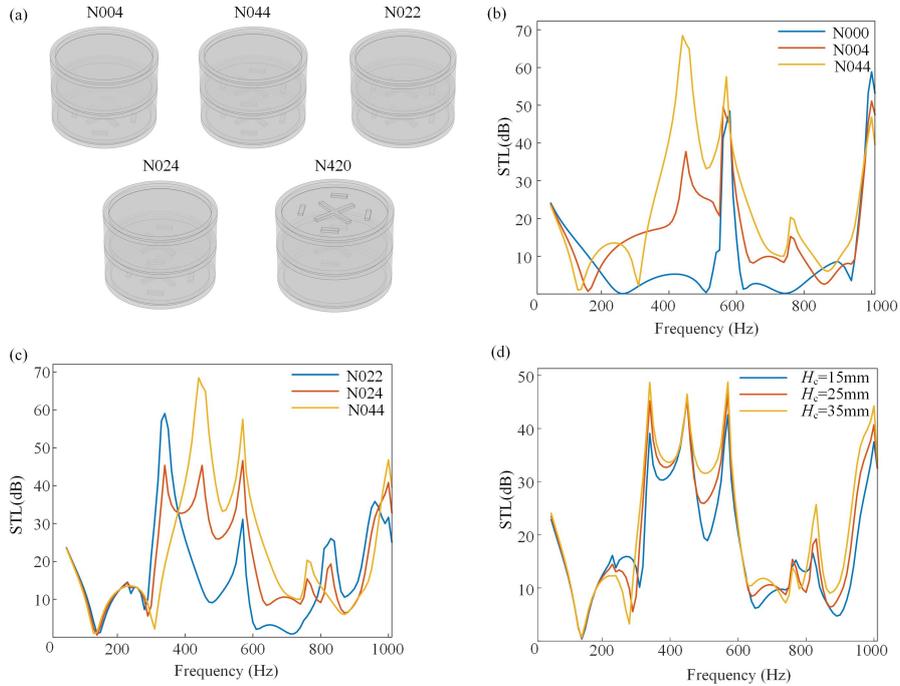

**Fig. 6.** (a) Different simulation structures; (b) comparison of the STLs for N000, N004 and N044; (c) comparison of the STLs for N022, N024 and N044; (d) impact of cavity thickness Hc on STL.

## 4. Sound insulation optimization

4.1 Orthogonal experiment and orthogonal list

Orthogonal experiment is a scientific method based on probability theory, mathematical statistics, professional knowledge and practical experience[26]. It uses standardized orthogonal tables to design and analyze the experimental results, reducing the number of experiments and determining the influencing factors. This method effectively improves experimental efficiency, saving time and resources. The characteristic values that need to be examined in the experiment, referred to as indicators, correspond to the experimental objectives. The factors that are examined in the experiment and may have an impact on the indicators are referred to as factors, typically represented by uppercase English letters such as A, B, C, etc., with each letter representing a factor. The selected states and conditions in which the factors are positioned in the experiment are referred to as levels. In order to accurately assess the sound insulation performance of the MAM cell, we take into account both the sound insulation bandwidth and the STL peak, and calculate the relative bandwidth ($\delta$). The relative bandwidth is determined by selecting a value that exceeds 30 dB near the STL peak. $\delta$ is designated as an inspection index, and its calculation is as follows:

$$\delta = (f_e - f_s) / f_s \tag{6}$$

where $f_s$ and $f_e$ are the start and stop frequencies, respectively. The prestressed force, membrane thickness and resonator width are designated as Factor A, Factor B and Factor C, respectively. Each factor is divided into 5 levels, as indicated in Table 2. Subsequently, an appropriate orthogonal array is chosen for the design, denoted as $L_n(t^c)$, where L represents the code of the orthogonal array, subscript n is the number of experiments, t denotes the factor levels, and superscript c signifies the maximum number of factors. In this case, the $L_{25}(5^6)$ orthogonal array is selected to perform 25 simulation calculations.

**Table 2.** The factor levels

| Level | Factor A | Factor B | Factor C |
|---|---|---|---|
| | $d_2$/mm | F/MPa | H/mm |
| 1 | 2 | 1.5 | 0.15 |
| 2 | 3 | 2 | 0.2 |
| 3 | 4 | 2.5 | 0.25 |
| 4 | 5 | 3 | 0.3 |
| 5 | 6 | 3.5 | 0.35 |

The range analysis method is applied to the results of 25 simulation calculations. The sum of the level data for each factor is denoted as $K_i$, and $\overline{K_i}$ represents the arithmetic mean of $K_i$, which is given by

$$\overline{K_i} = K_i/5 = (\sum_{i=1}^{5}\delta_i)/5 \tag{7}$$

where $\delta_i$ is the data of $i$ level of a factor, where $i$ = 1, 2, 3, 4 and 5. The range $R$ is the magnitude of variation for each factor as the data changes, indicating the degree of influence of each factor on the experimental results, which is given by

$$R = \max(\overline{K_i}) - \min(\overline{K_i}) \tag{8}$$

The calculated results are shown in Table 3. It is found that the variation in resonator width has the greatest impact on the sound insulation performance of the MAM cell, followed by membrane thickness and then prestressed force. The optimal solution is obtained by selecting the combination with the largest $K_i$ for each factor (A6, B6 and C2). The optimal combination is not present in the orthogonal table. By calculating the $\delta$ value of 1.2324, it is found to be greater than that of the experimental combinations. Through confirmatory experiments, the correctness of the orthogonal experimental results has been demonstrated.

**Table 3.** The results of the orthogonal experiment

| Test | $d_2$/mm | F/MPa | H/mm | $\delta$ |
|---|---|---|---|---|
| 1 | 2 | 1.5 | 0.15 | 0.3098 |
| 2 | 2 | 2 | 0.2 | 0.5728 |
| 3 | 2 | 2.5 | 0.25 | 0.5902 |
| 4 | 2 | 3 | 0.3 | 0.6137 |
| 5 | 2 | 3.5 | 0.35 | 0.6688 |
| 6 | 3 | 1.5 | 0.2 | 0.7174 |
| 7 | 3 | 2 | 0.25 | 0.8531 |
| 8 | 3 | 2.5 | 0.3 | 0.8130 |
| 9 | 3 | 3 | 0.35 | 0.7222 |
| 10 | 3 | 3.5 | 0.15 | 0.5712 |
| 11 | 4 | 1.5 | 0.25 | 0.8508 |
| 12 | 4 | 2 | 0.3 | 0.8786 |
| 13 | 4 | 2.5 | 0.35 | 0.8595 |
| 14 | 4 | 3 | 0.15 | 0.5750 |
| 15 | 4 | 3.5 | 0.2 | 1.0955 |
| 16 | 5 | 1.5 | 0.3 | 0.8443 |
| 17 | 5 | 2 | 0.35 | 0.8909 |
| 18 | 5 | 2.5 | 0.15 | 0.6346 |
| 19 | 5 | 3 | 0.2 | 1.2087 |
| 20 | 5 | 3.5 | 0.25 | 1.1928 |
| 21 | 6 | 1.5 | 0.35 | 0.8412 |
| 22 | 6 | 2 | 0.15 | 0.6109 |
| 23 | 6 | 2.5 | 0.2 | 1.1475 |
| 24 | 6 | 3 | 0.25 | 1.1435 |
| 25 | 6 | 3.5 | 0.3 | 1.1406 |
| $\overline{K_1}$ | 0.551 | 0.713 | 0.540 | |
| $\overline{K_2}$ | 0.735 | 0.761 | 0.948 | |
| $\overline{K_3}$ | 0.852 | 0.809 | 0.926 | |
| $\overline{K_4}$ | 0.954 | 0.853 | 0.858 | |
| $\overline{K_5}$ | 0.977 | 0.934 | 0.797 | |
| $R$ | 0.426 | 0.221 | 0.408 | |
| Optimum | 6 | 3.5 | 0.2 | 1.2324 |

4.2 Result

The comparison of STL between the optimal combination and the combination 024 is shown in Fig. 7. The STL values of the optimal combination are consistently higher than those of the combination 024 in the frequency range of 461.1-1038.5 Hz, and the sound insulation bandwidth exceeding 30 dB has significantly increased. It is indicated that through orthogonal experiments, multi-factor analysis can be conducted on MAM, leading to structural optimization and improvement in STL, as well as widening the sound insulation bandwidth.

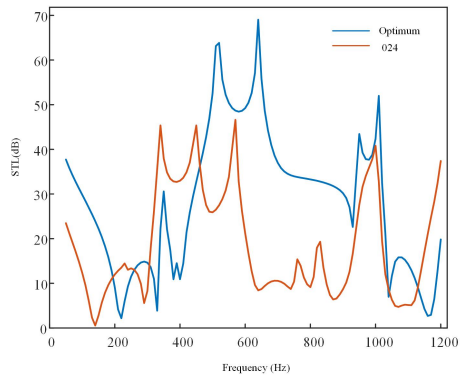

**Fig. 7.** Comparison of the STLs for optimum and N024.

## 5. Conclusion

In summary, this paper presents the design of a three-layer MAM structure and explores the sound insulation principle of MAM through the analysis of eigenmodes. It is found that incorporating different structures in each layer contributes to improving the overall sound insulation performance while maintaining lightweight characteristics. By utilizing orthogonal experiments, the main factors affecting sound insulation performance are identified, and the optimal combination of multiple parameters is determined, resulting in improved STL and expanded sound insulation bandwidth of the structure. This method can be applied to the optimization of other acoustic MAM metamaterials.

**Acknowledgments**

Project supported by National Natural Science Foundation of China (Grant No. 61571222), Natural Science Foundation of Jiangsu Province (Grant No. BK20210541), Natural Science Foundation of the Jiangsu Higher Education Institutions of China (Grant No. 21KJB140003) and Six Talent Peaks Project of Jiangsu Province, China.

**Data availabilityAVAILABILITY**

The data that support the findings of this study are available from the corresponding author upon reasonable request.